\documentclass[twocolumn,showpacs,preprintnumbers,amsmath,amssymb,superscriptaddress]{revtex4}

\usepackage{graphicx}% Include figure files
\usepackage{dcolumn}% Align table columns on decimal point
\usepackage{bm}% bold math

\let\a=\alpha \let\b=\beta  \let\g=\gamma     \let\d=\delta  \let\e=\varepsilon
  \let\h=\eta       \let\l=\lambda
\let\m=\mu    \let\n=\nu                  
\let\s=\sigma         
   \let\o=\omega 
 \let\D=\Delta

\def\\{\hfill\break} \let\==\equiv

\let\io=\infty 

\let\0=\noindent

\def\ie{\hbox{\it i.e.\ }}
\let\dpr=\partial

\def\tende#1{\,\vtop{\ialign{##\crcr\rightarrowfill\crcr
 \noalign{\kern-1pt\nointerlineskip}
 \hskip3.pt${\scriptstyle #1}$\hskip3.pt\crcr}}\,}
\def\otto{\,{\kern-1.truept\leftarrow\kern-5.truept\to\kern-1.truept}\,}

\def\T#1{{#1_{\kern-3pt\lower7pt\hbox{$\widetilde{}$}}\kern3pt}}
\def\VVV#1{{\underline #1}_{\kern-3pt
\lower7pt\hbox{$\widetilde{}$}}\kern3pt\,}
\def\W#1{#1_{\kern-3pt\lower7.5pt\hbox{$\widetilde{}$}}\kern2pt\,}

\def\lis{\overline}

\def\indica{\leaders \hbox to 0.5cm{\hss.\hss}\hfill}
\def\guida{\leaders\hbox to 1em{\hss.\hss}\hfill}
\mathchardef\oo= "0521

\def\xx{{\bf x}}
\def\yy{{\bf y}}\def\kk{{\bf k}}

\def\oo{{\underline \omega}}

\def\qed{\raise1pt\hbox{\vrule height5pt width5pt depth0pt}}

\def\indic{\hbox{\raise-2pt \hbox{\indbf 1}}}

\begin{document}

\preprint{APS/123-QED}

\title{Anomalous critical exponents in
the anisotropic Ashkin--Teller model}

\author{A. Giuliani}
%\email{alessandro.giuliani@roma1.infn.it}
\affiliation{%
Dipartimento di Fisica, Universit\'a di Roma ``La Sapienza'',\\
P.zzale A. Moro 2, 00185 Roma, Italia}
% \altaffiliation[Also at ]{INFN, Sezione di Roma 1}
 
\author{V. Mastropietro}%
%\email{mastropi@axp.mat.uniroma2.it}
\affiliation{%
Dipartimento di Matematica, Universit\'a di Roma ``Tor Vergata'',\\
Viale della Ricerca Scientifica 00133 Roma, Italia}

\date{\today}% It is always \today, today,

\begin{abstract}
We perform a rigorous computation 
of the specific heat of the Ashkin-Teller
model in the case of small interaction
and we explain how the universality--nonuniversality
crossover is realized when the isotropic limit is reached.
We prove that, even in the region where universality for the specific heat 
holds, anomalous critical exponents appear: for instance
we predict the existence 
of a previously unknown anomalous 
exponent, continuously varying with the strength
of the interaction, describing how the difference 
between the critical temperatures
rescales with the anisotropy parameter.
\end{abstract}

\maketitle

More than half a century ago 
Ashkin and Teller \cite{[AT]} introduced their model
as a generalization of the Ising model to a four component 
system. It describes a bidimensional lattice, each site of which 
is occupied by one of four kinds of atoms: $A$, $B$, $C$, $D$.
Two neighbouring atoms interact with an energy: $\e_0$ for
$AA$, $BB$, $CC$, $DD$; 
$\e_1$ for $AB$, $CD$; $\e_2$ for $AC$, $BD$; and $\e_3$ for 
$AD, BC$. Fan \cite{[F]} has shown that the AT model can be also written
in terms of Ising variables $\s^{(1)}_\xx=\pm 1, \s^{(2)}_\xx=\pm 1$
located at each site of the lattice; its Hamiltonian
can be written, if $\xx,\yy$ are
nearest neighbor sites as:
\begin{equation}
H^{AT}=\sum_{\xx, \yy} J^{(1)}
\s^{(1)}_\xx\s^{(1)}_{\yy}+
J^{(2)}
\s^{(2)}_\xx\s^{(2)}_{\yy}
+\l\s^{(1)}_\xx\s^{(1)}_{\yy}
\s^{(2)}_\xx\s^{(2)}_{\yy}\label{1}\end{equation}
with $J^{(1)}=\b(\e_0+\e_1-\e_2-\e_3)/4$,
$J^{(2)}=\b(\e_0+\e_2-\e_1-\e_3)/4$,
$\l=\b(\e_0+\e_3-\e_1-\e_2)/4$ and $\b$ is the inverse temperature.
The AT model is then equivalent to two Ising models
coupled by an interaction quartic in the spins; 
the case in which the two Ising subsystems are 
identical $J^{(1)}=J^{(2)}$ is called
{\it isotropic}, the opposite case {\it anisotropic}.
When the coupling $\l$ is $=0$, 
Ashkin--Teller (AT) reduces to two independent Ising models
and it has of course {\it two} critical temperatures if $J^{(1)}\not= J^{(2)}$.
 
Layers of atoms and molecules adsorbed on clean surfaces,
like submonolayers of Se adsorbed on Ni, are believed
to constitute physical realizations
of the AT model \cite{[Bak],[BE],[DR]}; theoretical
results on it can explain the phase diagrams 
of such systems, experimentally obtained 
by means of electron diffraction techniques.
As for the Ising model, the importance of AT is also in providing
a conceptual laboratory in which the higly non trivial
phenomenon of phase transitions
can be understood quantitatively in a relatively manegeable
model; in particular it has attracted great theoretical interest 
because is a simple and non trivial generalization of the Ising
and four-state Potts models.

Contrary to many $2d$ models
in statistical mechanics like the Ising, the 6 or
the 8 vertex models \cite{[Ba]}, in which remarkable exact
solutions give us very detailed informations
about the behaviour of thermodinamical functions,
there are no exact results on the AT
model except for the trivial 
$\l=0$ case. It is believed \cite{[WL]} that the AT
has {\it two} critical temperatures
for $J^{(1)}\not= J^{(2)}$ which coincide
at the isotropic point $J^{(1)}= J^{(2)}$.
Moreover it was conjectured by Kadanoff \cite{[K]}
and Baxter \cite{[Ba]} that 
the critical properties in the anisotropic 
and in the isotropic case are completely different; in
the first case the critical behaviour should be described
in terms of {\it universal} critical indices (identical
to those of the $2d$ Ising model)
while in the isotropic case the critical behaviour should be 
{\it nonuniversal} and described in terms
of indexes which are non trivial functions of $\l$.
In other words, the AT model should exhibit a 
{\it universal--nonuniversal} crossover when
the isotropic point is reached.

Evidence for the validity of nonuniversal behaviour
in the isotropic case was given in \cite{[PB]}
(using second order renormalization group arguments)
and in \cite{[LP],[N]} (by a heuristic mapping 
into the massive Luttinger model describing 
one dimensional interacting fermions
in the continuum). The anisotropic case was
studied numerically by 
Migdal--Kadanoff Renormalization Group \cite{[DR]},
Monte Carlo Renormalization group \cite{[Be]},
finite size scaling \cite{[Bad]}; such results give evidence of
the fact that, far away from the isotropic point, 
AT has two critical points and belongs to the same universality class
of the Ising model but give essentially no informations
on the critical behaviour when the anisotropy
is small.

In this Letter we present
a {\it rigorous} derivation of the specific
heat for the AT model, valid for small interaction $\l$ 
and any anisotropy.
We find indeed that 
in the anisotropic case the specific heat is singular in correspondence
of two critical temperatures, and the divergence
is {\it logarithmic} as in the Ising model,
in agreement with universality hypothesis. 
Nevertheless even in the region where universality
holds, {\it anomalous}
critical exponents appear; for instance 
the difference between the two critical temperatures
rescale with the anisotropy parameter with
a nonuniversal critical exponent. The presence
of such critical exponents also in the universality region
clarify 
how the universal--nonuniversal crossover is realized
when the isotropic limit is reached. 

Such results 
are found by the new methods introduced in \cite{[PS]} and \cite{[M]}
to study $2d$ statistical mechanics models which can be considered 
perturbations of the Ising model. These methods 
take advantage from the fact that such systems
can be exactly mapped in models of weakly interacting 
relativistic fermions in $d=1+1$ 
on a lattice. The mapping was known
since long time (see Ref. \cite{[LSM],[LP],[ID],[S]});
however in recent 
years a great progress has been achieved in the 
evaluation of Grassmann integrals of interacting models,
in the context of quantum field theory
and solid state physics
(see Ref. \cite{[BG],[FMRT],[GeM]}),
and one can take advantage of this new technology 
to get informations about $2d$ statistical mechanics models.
This provides the only method to get rigorous quantitative 
informations on the critical properties of such systems if an
exact solution is lacking, as in the present case. 
The algorithm is based 
on multiscale analysis and 
allows us to prove
{\it convergence} of the expansion for the energy--energy
correlation functions and for the specific heat {\it up to the
critical temperature}; essential ingredients of our analysis are
cancellations due to anticommutativity
of fermionic variables and approximate {\it Ward identities} \cite{[BM]}, guaranteeing that
the flow of the effective
coupling constants is not diverging in the infrared region.
We stress that our method applies to a large class
of perturbations of the 2d Ising model, and for sake of definiteness
we restrict our analysis to AT. 

In order to present our result, we find convenient 
to introduce the variables 
\begin{equation}
t={t^{(1)}+t^{(2)}\over 2}
\;,\quad u={t^{(1)}-t^{(2)}\over 2}\label{3}
\end{equation}
with $t^{(j)}=\tanh J^{(j)}$, $j=1,2$. 
The parameter $t$ has the role of a {\it reduced temperature}
and $u$ measures the {\it anisotropy}
of the system. We shall consider the free
energy or the specific heat as functions of $t,u,\l$.
When $\l=0$ the specific heat $C_v$
can be immediately computed from the Ising model
exact solution; $C_v$ is diverging at 
$t=t_c^\pm=\sqrt2-1\pm|u|$ and near the 
the critical temperatures the specific heat
shows a logarithmic divergence:
$C_v\simeq -C\log|t-t_c^\pm|$, where $C>0$.
If the anisotropy is strong 
the two Ising subsystems 
have very different critical temperatures,
hence one can expect that if one system is almost critical
the second one will be out 
of criticality; then mean field arguments based
on the fact that two Ising are
coupled by a density-density interaction
suggest that the effect of the coupling
is to change at most the value
of the critical temperatures. 
On the other hand if the anisotropy is small
the two system will become critical almost at the same 
temperature and the properties of the system could change drastically.

Our main result is the following theorem; the 
detailed proof can be found in \cite{[GM],[M]}.
\vskip.5cm
{\bf Theorem.}\ {\it For $\l$ small enough 
the AT model admits 
two critical points of the form:
$$t_c^\pm(\l,u)=\sqrt2-1+\n(\l)\pm |u|^{1+\h}(1+\d(\l,u))\;.$$
Here $\n$ and $\d$ are $O(\l)$ corrections and $\h=-b\l+O(\l^2)$
with $b>0$.
If $t\not=t_c^\pm$
the free energy of the model is analytic in $\l,t,u$ and 
the specific heat $C_v$ is equal to:
\begin{equation}
-F_1\D^{2\h_c}\log{|t-t_c^-|\cdot|t-t_c^+|\over \D^2}+F_2{1-\D^{2\h_c}
\over \h_c}+F_3\;,\end{equation}
where: $2\D^2=(t-t_c^-)^2+(t-t_c^+)^2$; $\h_c=a\l+O(\l^2)$, $a\not=0$;
and $F_1$, $F_2$, $F_3$ are functions of $t,u,\l$, bounded above and below
by $O(1)$ constants.}
\\

\01) First note that the location of the critical points is
dramatically changed 
by the interaction. The difference of
the interacting critical temperatures normalized with the free one
$G(\l,u)\=(t_c^+(\l,u)-t_c^-(\l,u))/(t_c^+(0,u)-t_c^-(0,u))$
rescales with the anisotropy parameter as
a power law $\sim|u|^{\h}$, and in the limit $u\to 0$
it vanishes or diverges, depending on the sign of $\l$ 
(this is because $\h=-b\l+O(\l^2)$, 
with $b>0$). In Fig. 1 we plot the qualitative behaviour
of $G(\l,u)$ as a function of $u$, for two different values of $\l$
(\ie we plot the function $u^\h$, with $\h=0.3,-0.3$ respectively).

\begin{figure}[ht]
\centering
\includegraphics[width=.52\textwidth,angle=0]{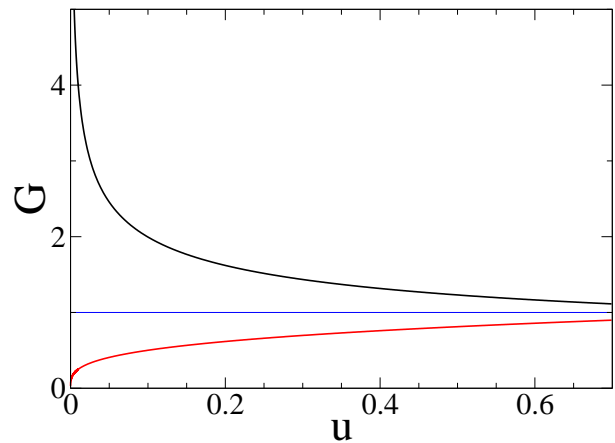}
\caption{\label{fig:epsart} The behaviour of the difference $G$
between the interacting critical temperatures normalized
to the free one, for two different values of $\l$;
depending on the sign of the interaction, it diverges
or vanishes in the isotropic limit. 
}
\end{figure}

As far as we know, the existence of
the critical index $\h(\l)$ was not known in the literature,
even at a heuristic level.\\

\02) There is universality for the 
specific heat, in the sense that it diverges logarithmically
at the critical points, as in the Ising model.
However the coefficient of the log
is {\it anomalous}: in fact if $t$ is near to one of the critical temperatures 
$\D\simeq\sqrt{2}|u|^{1+\h}$ so that the coefficient in front of the logarithm
behaves like $\sim |u|^{2(1+\h)\h_c}$, with 
$\h_c$ a new anomalous exponent $O(\l)$; 
in particular it is vanishing or diverging as 
$u\to 0$ depending on the sign of $\l$.
We can say that the system shows an {\it anomalous universality}
which is a sort a new paradigmatic behaviour: 
the singularity at the critical 
points is described in terms of universal critical indexes
nevetheless in the isotropic limit $u\to 0$, 
some quantities, like the difference
of the critical temperatures and the constant in front of the logarithm 
in the specific heat, scale with anomalous critical indexes, and they
vanish or diverge, depending on the sign of $\l$.\\

\03) Eq(3) clarifies how the universality--nonuniversality
crossover is realized as $u\to 0$.
When $u\not=0$ only the first term in eq(3) can be log--singular
in correspondence of the two critical points; 
however
the logarithmic term dominates on the second one only if $t$ varies 
inside an extremely small region $O(|u|^{1+\h}e^{-c/|\l|})$
around the 
critical points (here $c$ is a positive $O(1)$ constant). 
Outside such region the power law behaviour 
corresponding to the second addend dominates. When $u\to 0$ one
recovers the power law decay found in the isotropic case
$$C_v\simeq F_2{1-|t-t_c|^{2\h_c}\over\h_c}$$
In Fig. 2
we plot the qualitative behaviour of $C_v$ 
as a function of $t$. 
The three graphs are 
plots of eq(3), with $F_1=F_2=1$, $F_3=0$, 
$u=0.01$, $\h=\h_c=0.1,0,-0.1$ respectively; the central curve 
corresponds to the case $\h=0$, the upper one to $\h<0$ and 
the lower to $\h>0$.

\begin{figure}[ht]
\centering
\includegraphics[width=.51\textwidth,angle=0]{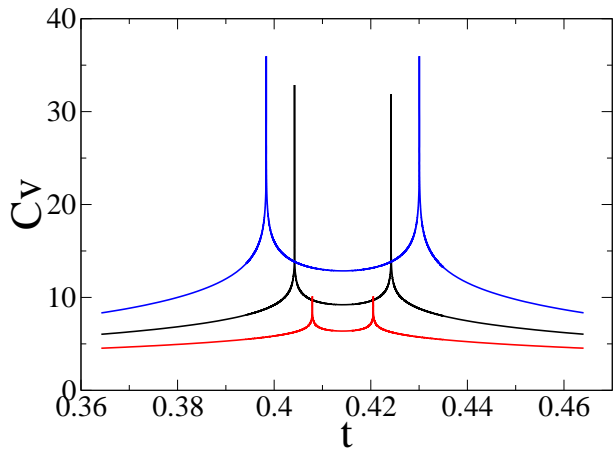}
\caption{\label{fig:epsart} The behaviour
of the specific heat $C_v$ 
for three different values of $\l$, showing the log--singularities
at the critical points;
in the isotropic limit the two critical points tend tocoincide,
the lower curve becomes continuous while the upper 
develops a power law divergence.}
\end{figure}

We now sketch the proof of the
above Theorem (for a detailed proof we refer to
\cite{[GM],[M]}). We start from the well known 
representation of the Ising model 
free energy in terms of a sum of {\it Pfaffians} \cite{[MW]}
which can be equivalently written (see Ref.
\cite{[ID],[S]}) 
as {\it Grassmann functional integrals}, formally describing
massive non interacting Majorana fermions 
$\psi,\lis\psi$ on a lattice with action
\begin{eqnarray}
&&\sum_\xx{t\over 4}\Big[
\psi_\xx(\partial_1-i\partial_0)\psi_\xx+
\lis\psi_\xx(\partial_1+i\partial_0)\lis\psi_\xx-\\
&&\qquad-2
i\lis\psi_\xx(\partial_1+\dpr_0)\psi_\xx\Big]
+i(\sqrt{2}-1-t)\lis\psi_\xx\psi_\xx\;,\nonumber\end{eqnarray}
where $\dpr_j$ are discrete derivatives; criticality
corresponds to the massless case. If
$\l=0$ the free energy and specific
heat of the AT model
can be written as sum of Grassmann integrals describing {\it two}
kinds of Majorana fields, with masses $m^{(1)}=t^{(1)}-\sqrt2+1$
and $m^{(2)}=t^{(2)}-\sqrt2+1$.

If $\l\not=0$ again the free energy and the specific heat
can be written as Grassmann integrals, but the Majorana
fields are {\it interacting} with a short range potential.
By performing a suitable change
of variables and integrating out the ultraviolet
degrees of freedom, the effective action can  
can be written as
\begin{eqnarray}
&&Z_1\sum_{\xx,\o,\a}\Big[
\psi^+_{\o,\xx}(\dpr_1-i\o\dpr_0)\psi^-_{\o,\xx}
-i\o\s_1\psi^+_{\o,\xx}\psi^-_{-\o,\xx}+\nonumber\\
&&+i\o\m_1\psi^\a_{\o,\xx}
\psi^\a_{-\o,-\xx}+
\l_1\psi^+_{1,\xx}\psi^-_{1,\xx}\psi^+_{-1,\xx}\psi^-_{-1,\xx}\Big] +
{\cal W}_1\nonumber
\end{eqnarray}
where $\a=\pm$ is a {\it creation--annihilation} index
and $\o=\pm 1$ is a {\it quasi--particle} index.
$\s_1$ and $\m_1$ have the role of two {\it masses}
and it holds
$\s_1=O(t-\sqrt2+1)+O(\l)$, $\m_1=O(u)$. ${\cal W}_1$
is a sum of monomials of $\psi$ of arbitrary
order, with kernels which
are {\it analytic functions} 
of $\l_1$; analyticity is a very nontrivial property
obtained exploiting anticommutativity
properties of Grassman variables via {\it Gram inequality}
for determinants. The 
$\psi^\pm$ are {\it Dirac} fields,
which are
combinations of the Majorana variables $\psi^{(j)},\lis\psi^{(j)}$, $j=1,2$,
associated with the two Ising subsystems.

One can compute the partition function by expanding 
the exponential of the action 
in Taylor series in $\l$ and naively
integrating term by term the Grassmann monomials, using the Wick rule;
however such a procedure gives poor bounds for the coefficients
of this series that, in the thermodynamic
limit, can converge 
only far from the critical points.

In order to study the 
critical behaviour of the system
we perform a multiscale analysis 
involving non trivial resummations of the perturbative series.
The first step is to decompose the 
propagator $\hat g(\kk)$ as a sum of propagators more and more singular 
in the infrared 
region, labeled by an integer $h\le 1$, so that $\hat g(\kk)=
\sum_{h=-\io}^1\hat g^{(h)}(\kk)$,
$\hat g^{(h)}(\kk)\sim\g^{-h}$. 
We compute the Grassmann integrals defining the partition function
by iteratively integrating
the propagators
$\hat g^{(1)},\hat g^{(0)},\ldots$
After each integration step we rewrite the partition function in a way 
similar to the last equation,
with $Z_h,\s_h,\m_h,\l_h,{\cal W}_h$
replacing $Z_1,\s_1,\m_1,\l_1,{\cal W}_1$, in particular 
the masses 
and the wave function renormalization are modified;
the structure of 
the action is preserved because of
symmetry properties; moreover
${\cal W}_h$ is shown to be a sum of monomials of $\psi$ of arbitrary
order, with kernels decaying in real space on scale $\g^{-h}$, which
are {\it analytic functions} 
of $\{\l_h,\ldots,\l_1\}$,
if $\l_k$ are small enough, $k\ge h$, and $|\s_k|\g^{-k},|\m_k|\g^{-k}\le 1$;
again analyticity follows from Gram--Hadamard type of bounds.

All the above construction is based 
on the crucial property that the effective interaction
at each scale does not increase
$|\l_h|\le 2|\l|$; such property is a consequence of the validity
of some non perturbative {\it approximate Ward identities} 
\cite{[BM]}; ``approximate'' refers to the fact that,
because of the presence of masses and of an ultraviolet cutoff,
the Ward identities are different from the usual formal
ones; the error terms are shown to be small, in a suitable sense.
For $\s_h,\m_h,Z_h$, we find that, under the iterations, they 
evolve as: $\s_h\simeq\s_1 \g^{b_2\l h}$, $\m_h\simeq \m_1 \g^{-b_2\l h}$, 
$Z_h\simeq\g^{-b_1\l^2h}$, with $b_1, b_2$ explicitely computable
in terms of a convergent power series.

We perform the iterative integration described above up to a
scale $h^*_1$ such that $(|\s_{h^*_1}|+|\m_{h^*_1}|)\g^{-h^*_1}=O(1)$.
For scales lower than $h^*_1$ we return to the description in terms
of the original Majorana fermions
$\psi^{(1, \le h^*_1)}$, $\psi^{(2, \le h^*_1)}$
associated with the two Ising subsystems.
One of the two fields (say $\psi^{(1,\le h^*_1)}$) 
is massive on scale $h^*_1$
(so that the Ising subsystem with $j=1$ is ``far from criticality'' 
on the same scale); then 
we can integrate
the massive Majorana
field $\psi^{(1,\le h^*_1)}$ without any further 
multiscale analysis, obtaining
an effective theory
of a single Majorana field with mass $|\s_{h^*_1}|-|\m_{h^*_1}|$, which 
can be arbitrarly small; this is equivalent
to say that on scale $h^*_1$
we have an effective description of the system as a single 
perturbed Ising model
with {\it anomalous} parameters near criticality.
The integration of the scales $\le h^*_1$ is performed again by
a multiscale decomposition similar to the one just described; an
important feature is however that there are no more quartic
marginal terms, because the anticommutativity
of Grassmann variables forbids local quartic monomials
of a single Majorana fermion. 
Criticality 
is found when the effective mass on scale $-\io$ is vanishing;
the values of $t,u$ for which this happens
are found by solving a non trivial implicit function problem.

Technically it is an interesting
feature of this problem that there are two regimes in which the
system must be described in terms of different fields:
a first one in which the natural variables are Dirac Grassmann variables,
and a second one in which they are Majorana; the scale $h^*_1$
separating the two
regimes is dynamically generated by the iterations. 
In the first regime
the two entangled Ising subsystems are 
undistinguishable,
the natural description is in terms of Dirac variables and
the effective interaction is marginal; in the integration
of such scales nonuniversal indexes appear. In the second region
the two Ising subsystems really look different, one appear to be 
(almost) at criticality and the other far from criticality on 
the same scale; the 
parameters of the two subsystems are deeply changed 
(in an anomalous way) by the previous integration;
in this region the effective interaction is irrelevant.

In conclusion we have presented some new rigorous results
on the critical behaviour in the Ashkin--Teller model, 
for weak coupling and 
any value of the anisotropy. Via multiscale integration 
methods we have 
computed the specific heat and the location of the critical 
temperatures
in terms of {\it convergent} power series and we have predicted 
the existence of an unknown critical exponent describing the scaling 
of the gap between the critical temperatures in the isotropic limit. 
Moreover we gave a detailed description of the crossover between the
universal critical behaviour holding in the anisotropic case 
and the anomalous nonuniversal behaviour holding in the isotropic limit.

\end{document}